\documentclass{pasj00}

\draft

\usepackage{times}

\begin{document}
\SetRunningHead
{S. Okumura et al.}
{Physical relation of source I to IRc2 in the Orion KL Region}
\Received{2011/01/07}%{yyyy/mm/dd}
\Accepted{2011/04/21}%{yyyy/mm/dd}
\Published{$\langle$publication date$\rangle$}

\title{Physical Relation of Source I to IRc2 in the Orion KL Region
\thanks{Based on data collected at Subaru Telescope and obtained 
from the SMOKA science archive, which is operated by the Astronomy 
Data Center, National Astronomical Observatory of Japan.}}

\author{Shin-ichiro \textsc{Okumura},\altaffilmark{1,2}
			Takuya \textsc{Yamashita},\altaffilmark{3}
			Shigeyuki \textsc{Sako},\altaffilmark{4}
			Takashi \textsc{Miyata},\altaffilmark{4}\\
			Mitsuhiko \textsc{Honda},\altaffilmark{5}
			Hirokazu \textsc{Kataza}\altaffilmark{6},
			and
			Yoshiko K. \textsc{Okamoto}\altaffilmark{7}
			}
\altaffiltext{1}{Bisei Spaceguard Center, 
Japan Spaceguard Association,\\
				1716--3 Okura, Bisei-cho, 
				Ibara-shi, Okayama 714--1411}
\altaffiltext{2}{Okayama Astrophysical Observatory, 
				National Astronomical Observatory of Japan, \\
				Kamogata-cho, Asakuchi-shi, Okayama 719--0232}
\altaffiltext{3}{National Astronomical Observatory of Japan, 
				2--21--1 Osawa, Mitaka-shi, Tokyo 181--8588}
\altaffiltext{4}{Institute of Astronomy, Graduate School of Science, 
				University of Tokyo, \\
				2--21--1 Osawa, Mitaka-shi, Tokyo 181--0015}
\altaffiltext{5}{Department of Information Science, 
				Kanagawa University, 
				\\2946 Tsuchiya, Hiratsuka-shi, Kanagawa 259--1293}
\altaffiltext{6}{Institute of Space and Astronautical Science, 
				Japan Aerospace Exploration Agency,\\
				3--1--1 Yoshinodai, Chuo-ku, 
				Sagamihara-shi, Kanagawa 252--5210}
\altaffiltext{7}{Faculty of Science, Ibaraki University, 
				2--1--1 Bunkyo, Mito-shi, Ibaraki 310--8512}

\email{okumura@spaceguard.or.jp}

\maketitle

\begin{abstract}
We present mid-infrared narrow-band images of 
the Orion BN/KL region, and $N$-band low-resolution spectra of 
IRc2 and the nearby radio source ``I.'' 
The distributions of the silicate absorption strength and 
the color temperature have been
revealed with a sub-arcsecond resolution. 
The detailed structure of the 7.8 $\micron$/12.4 $\micron$ 
color temperature distribution 
was resolved in the vicinity of IRc2.
A mid-infrared counterpart to source I has been detected 
as a large color temperature peak.
The color temperature distribution shows
an increasing gradient from IRc2 toward source I,
and no dominant temperature peak is seen at IRc2.
The spectral energy distribution of IRc2 could be
fitted by a two-temperature component model,
and the ``warmer component'' of the infrared emission from IRc2 
could be reproduced by scattering of radiation from source I.
IRc2 itself is not self-luminous,
but is illuminated and heated by an embedded luminous 
young stellar object located at source I.

\end{abstract}

\section{Introduction}
The Orion Molecular Cloud, at a distance of 
418 $\pm$ 6 pc \citep{vera},
is the closest and best-studied region of massive star formation.
After the discovery of BN  \citep{bn} and KL \citep{kl},
following infrared observations
revealed that the KL nebula splits up into a number of compact 
``IRc'' sources of different color temperatures \citep{rie73}.
The most luminous of these, IRc2, was thought to be the dominant
energy source for the KL complex \citep{dow81,wyn84}.

Radio observations have detected H${_2}$O masers 
\citep{hil72,sul73,gen77,mor77} and
SiO masers \citep{sny74, mor77, wri83}
near IRc2, thought to be a common central point at that time.
In addition, radio continuum point source ``I'' was found 
at the SiO maser position \citep{RSI1,RSI2}, and \citet{men95}
pointed out that source I is located precisely 
at the centroid of the SiO maser distribution.
Other observations also revealed
that source I and SiO masers are surrounded 
by the H${_2}$O ``shell'' masers 
\citep{gen89,wri90,gau98}.

More recent high-resolution radio and infrared observations 
clarified that the infrared position of IRc2 is offset 
from the radio position of source I 
\citep{gez92,dou93,men95}.
\citet{dou93} also resolved IRc2 into a number of components 
at 3.8 $\micron$,
suggesting the presence of several embedded stars.
\citet{gez98} estimated that the total infrared luminosity of IRc2 is
only $L\sim$1000$\LO$, approximately 2 orders of magnitude 
lower than the previously accepted value.
Therefore, 
it is generally accepted that IRc2 is not a very luminous single star 
which significantly contributes to the energetics of the KL complex. 

The nature of IRc2, heated internally or externally, is still unclear.
Intermediate cases were suggested, 
with some IRc2 components being self-luminous
and others not \citep{gez98,rob05}.
The external illuminating source has not been identified,
though source I is a prime candidate.
The relationship of source I to IRc2 also
still remains ambiguous.
In this paper, we present evidence of
a direct linkage between IRc2 and source I
with subarcsecond angular resolution mid-infrared data,
and interpret that IRc2 is illuminated externally by source I.

\section{Observations and Data Analysis}

\subsection{Observations and data acquisition}

Mid-infrared imaging and spectroscopic observations 
of the Orion BN/KL region
were made on 2002 January 2 and 2000 December 10, respectively, 
with the Cooled Mid-Infrared Camera and Spectrometer,
COMICS \citep{comics1,comics2} mounted on the 8.2-m SUBARU telescope.
The data were retrieved from the SUBARU archive system, 
SMOKA \citep{smoka}.
Imaging observations were made in nine bands 
from 7.8 to 24.8 $\micron$.
The pixel scale was \timeform{0".13},
and the total field of view was $\timeform{31"} \times \timeform{41"}$.
To cancel the high background radiation,
the secondary mirror chopping was used
at a frequency of 0.5 Hz with a \timeform{30"} throw and 
a direction to the position angle of \timeform{-90D}.
A nodding technique was not used in the present observations
because the object was bright enough compared to the fluctuation
level of the remaining sky pattern after chopping subtraction. 
During imaging observations,
$\alpha$ CMi and $\alpha$ Ori were observed  for flux calibration.
The full widths at half maximum (FWHMs) of the point spread function 
were measured to be \timeform{0".35}--\timeform{0".65} 
for $\alpha$ CMi and $\alpha$ Ori.
The parameters of our imaging observations are 
summarized in table \ref{tabobsim}.
$N$-band low resolution ($\lambda/\Delta\lambda$ $\sim$250)
spectroscopic observations were performed 
with the \timeform{0.33''}-wide, north-south slit.
The pixel scale along the slit was \timeform{0".165}.
The observations were carried out on three slit positions
in steps of \timeform{0.33''} from west to east,
on the eastern  part of IRc2 including source I.
During spectroscopic observations,
the slit position was confirmed by referring to 8.8 $\micron$
slit-viewer images taken simultaneously.
$\alpha$ Tau was observed and used 
for the correcting of the atmospheric absorption and flux calibration,
based on \citet{coh99}.
The parameters of our spectroscopic observations are 
summarized in table \ref{tabobssp}.

%%%%% Table 1 %%%%%
\begin{table*}
\caption{Parameters of the imaging observations on 2002 January 2}
\label{tabobsim}
\begin{center}
   \begin{tabular}{lcccc}
   \hline
      Object  & Filter 
     & Integ. Time & Air Mass & FWHM\\
	 & ($\micron$) & (sec) && (arcsec)\\
	 \hline
      BN/KL \dotfill & 7.8 ($\Delta\lambda$ = 0.7) &
	  	19.6 & 1.349--1.311&0.40\footnotemark[$*$] \\
		 & 8.8 ($\Delta\lambda$ = 0.8) 
		 & 39.8 & 1.285--1.264 &0.43\footnotemark[$*$] \\
		 & 9.7 ($\Delta\lambda$ = 0.9)
		 & 39.2 & 1.242--1.215 &0.44\footnotemark[$*$] \\
         & 10.5 ($\Delta\lambda$ = 1.0)
		 & 40.2 & 1.205--1.188 &0.46\footnotemark[$*$] \\
		 & 11.7 ($\Delta\lambda$ = 1.0)
		 & 39.7 & 1.180--1.165 &0.46\footnotemark[$*$] \\
		 & 12.4 ($\Delta\lambda$ = 1.2)
		 & 39.7 & 1.159--1.146&0.42\footnotemark[$*$]  \\
		 & 18.5 ($\Delta\lambda$ = 0.9)
		 & 4.8 & 1.396--1.451&0.66\footnotemark[$*$]  \\
		 & 20.8 ($\Delta\lambda$ = 0.9)
			& 7.9 & 1.484--1.509 &0.70\footnotemark[$*$] \\
		 & 24.8 ($\Delta\lambda$ = 1.9)
			& 3.8 & 1.524--1.559 &0.96\footnotemark[$*$] \\
	$\alpha$ CMi \dotfill & 7.8 ($\Delta\lambda$ = 0.7) & 1.0 
	& 1.284 & 0.39\\
	& 8.8 ($\Delta\lambda$ = 0.8) & 1.2 & 1.270& 0.35\\
	& 9.7 ($\Delta\lambda$ = 0.9) & 1.0 & 1.265 & 0.59\\
	& 10.5 ($\Delta\lambda$ = 1.0) & 1.0&1.260& 0.60\\
	& 11.7 ($\Delta\lambda$ = 1.0) & 1.0&1.240& 0.43\\
	& 12.4 ($\Delta\lambda$ = 1.2) & 1.0&1.244 &0.46\\
	$\alpha$ Ori \dotfill & 18.5 ($\Delta\lambda$ = 0.9) & 1.0 
	&1.338& 0.53\\
	& 20.8 ($\Delta\lambda$ = 0.9) & 1.0 &1.328&0.57\\
	& 24.8 ($\Delta\lambda$ = 1.9) & 1.0 &1.315&0.65 \\
	  \hline
	\multicolumn{1}{@{}l@{}}{
	\hbox to 100pt{\parbox{150mm}{\footnotesize
   \footnotemark[$*$] The image size measured for BN.}
	\hss
	 }
	 }
    \end{tabular}
  \end{center}
\end{table*}

%%%%% Table 2 %%%%%
\begin{table*}
\caption{Parameters of the spectroscopic observations on 
					2000 December 10}
\label{tabobssp}
\begin{center}
   \begin{tabular}{ccccc}
   \hline
      Object  & Integ. Time & Air Mass & FWHM at 8.8$\micron$& 
	  Resolution\\
	& (sec) & & (arcsec)&\\
	 \hline
     IRc2 \dotfill&  90.4\footnotemark[$*$] &
		  1.285--1.231 &0.63\footnotemark[$\dagger$] &250\\
	$\alpha$ Tau \dotfill & 4.0 & 1.076--1.071 &
	0.50\footnotemark[$\ddagger$]&250\\
	  \hline
    \multicolumn{1}{@{}l@{}}{
	\hbox to 100pt{\parbox{85mm}{\footnotesize
   \footnotemark[$*$] Total on-source integration time 
   for each slit position.\\
   \footnotemark[$\dagger$] Measured for BN 
   in the 8.8$\micron$ slit-viewer images.\\
   \footnotemark[$\ddagger$] Measured on the spectroscopic data.}
	\hss
	 }
	 }
	    \end{tabular}
  \end{center}
\end{table*}

\subsection{Data reduction}

The data were reduced by using IRAF
	\footnote{IRAF is distributed by the National Optical Astronomy 
	Observatories, which is operated by the Association of Universities
	for Research in Astronomy (AURA), Inc., under cooperative 
	agreement with the National Science Foundation.}
and our own reduction tools.
For imaging data,
after the standard chopping subtraction, 
the flat-fielding was carried out by using blank-sky images taken 
at the chopping-OFF (off-source) position.
All flat-divided images were adjusted for the position 
referring to BN,
with an accuracy of 0.1 pixel (corresponding to \timeform{0".01}),
and then were co-added to produce a final image of size 
$\sim$\timeform{28"}$\times$\timeform{36"}.
For the spectroscopic data, 
after the procedure of chopping subtraction,
the thermal spectra of the telescope cell-cover was used for flat-fielding.
The wavelength scale was calibrated by atmospheric emission lines,
and the distortion of the spectral image on the detector was corrected
based on the spectra of the standard stars.
The extracted spectra were divided by the spectrum of
$\alpha$ Tau, which was in advance divided by a blackbody curve at a
temperature of 4400 K.

\subsection{Astrometric calibration}
\label{sec:astrometry}

The astrometry has been determined with respect to the  
position of BN on the final co-added images.
To register the radio position of source I \citep{men95}
and the 3.8 $\micron$ near-infrared positions 
\citep{dou93} on our images,
we assumed that the radio, near-infrared, and mid-infrared centroids 
for BN were coincident,
while BN is reported to have its own proper motions.
\citet{pla95} formerly pointed out that BN and source I are moving 
apart at $\sim$\timeform{0.02"} yr$^{-1}$, and
\citet{rod05} presented the absolute proper motions of 
BN and source I.
\citet{gom05} determined the mean absolute motion of Orion,
and registered all absolute proper motions to 
the ``rest'' frame of the Orion Nebula.

In applying the registration, the positions of BN and source I 
on other wavelength images
were corrected to the positions on our images taken in 2002,
by referring to the relative proper motions to 
the rest frame of the Orion Nebula,
reported by \citet{gom05}:
BN has a proper motion of 11.14$\pm$1.35 mas yr$^{-1}$ 
to the northwest, while source I, 7.79$\pm$1.89 mas 
yr$^{-1}$ to the southeast.
In the registration, 
IRc2 is assumed to stand on the rest frame of the Orion Nebula.
The astrometric corrections for individual sources amount to 
\timeform{0".13} in their maximum.
The uncertainty in the registration, 
mainly caused by the proper motion,
is estimated to be $\leq$ 1 pixel in our images.

\section{Results}
\subsection{Morphology}

%%%%% Figure 1 %%%%%
\begin{figure*}
  \begin{center}
	\FigureFile(160mm,111mm){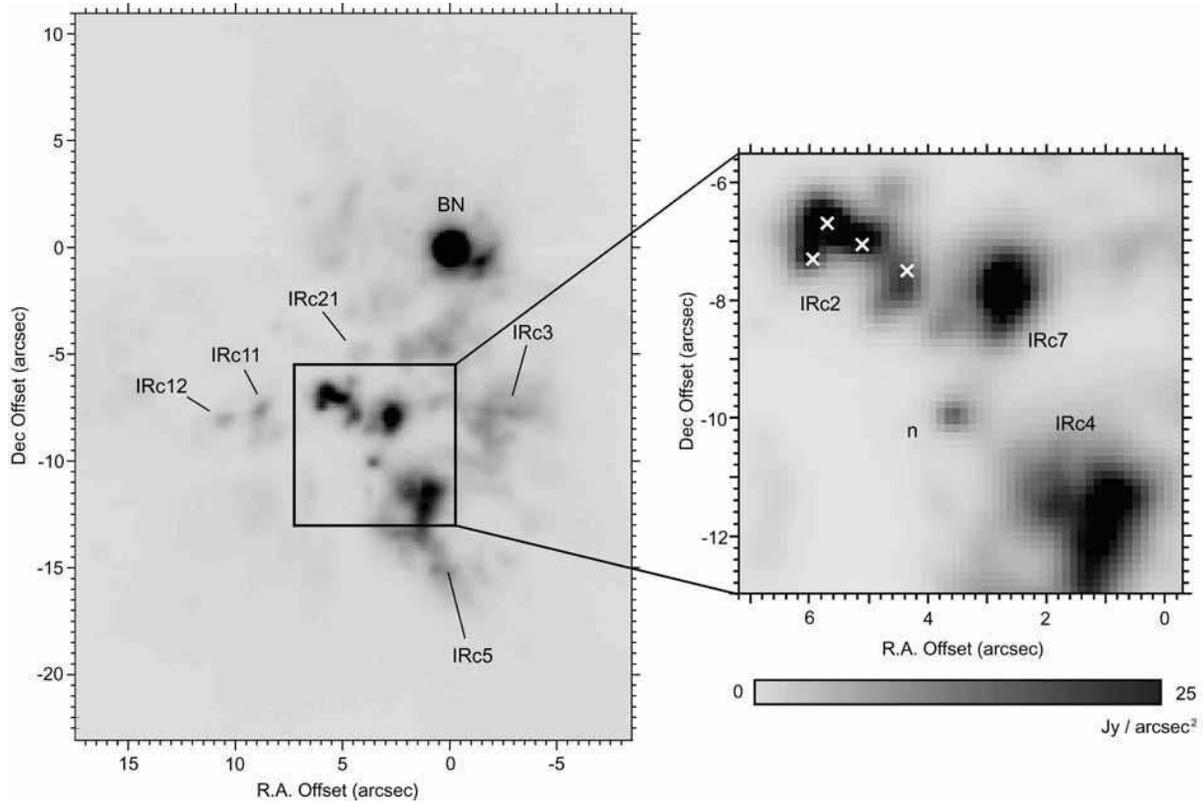} 
	\end{center} 
	\caption{12.4 $\micron$ image of the \timeform{26"}$\times$ 
	\timeform{34"} field of the BN/KL region (left) and
	the close-up image around IRc2 (right).
	Four crosses mark the positions of the 3.8 $\micron$
	IRc2-A, B, D, C, from left to right \citep{dou93}.
	Coordinates are centered on the BN object.
	}\label{12umimage} 
\end{figure*}

\begin{figure*}
  \begin{center}
		\FigureFile(100mm,131mm){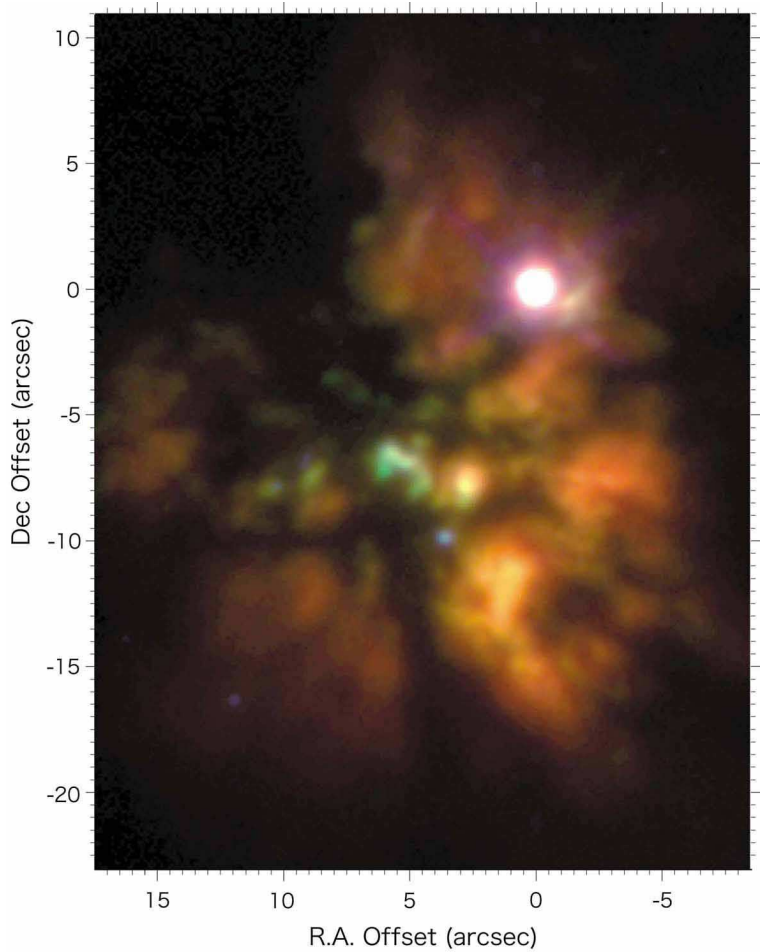}
	\end{center} 
	\caption{Three-color composite image of the BN/KL region 
	(\timeform{26"}$\times$\timeform{34"}): 8.7 $\micron$ (blue);
	12.4 $\micron$ (green); and 18.5 $\micron$ (red).
	}\label{3cimage} 
\end{figure*}

Figure \ref{12umimage} shows a 12.4 $\micron$ image
of the Orion BN/KL region (\timeform{26"}$\times$\timeform{34"}),
and figure \ref{3cimage} gives a three-color 
(8.7 $\micron$, 12.4 $\micron$, and 18.5 $\micron$)
composite image of the same region.
Our images agree well with the similar resolution mid-infrared data of
\citet{gre04} or \citet{shu04}, 
taken at the Keck-I telescope with the Long-Wavelength Spectrometer. 
In figure \ref{3cimage},
IR source n has a blue-color appearance, 
which represents a higher color temperature,
whereas diffuse emission appears mostly to be red,
which shows a lower color temperature. 
Several compact sources (IRc2, IRc12, IRc21, 
the southern part of IRc7, and the northern half of IRc11)
stand out as greenish appearance in the image.
No local peaks were detected at the source I position
[(R.A., Dec.)$_{\mathrm{offset}}$ 
= (\timeform{6.0"}, \timeform{-7.7"}) from BN], 
in any wavelength from 7.8 to 24.8 $\micron$.

%%%%% Figure 2 %%%%%

\subsection{Silicate absorption and color temperature distribution
     		around IRc2}
\label{sec:ext}

The strength of the 9.7 $\micron$ silicate absorption feature and 
the 7.8 $\micron$/12.4 $\micron$ color temperature
have been calculated for each pixel.
The procedure is as follows:

\begin{enumerate}

\item The first-order 9.7 $\micron$ silicate absorption factor,
$exp(\tau_{9.7}$), 
is approximated 
as the ratio of the observed 9.7 $\micron$ brightness to 
the 9.7 $\micron$ continuum,
 linearly interpolated between 7.8 and 12.4 $\micron$.

\item The 7.8 and the 12.4 $\micron$ brightness are corrected 
(dereddened) based on the first-order 9.7 $\micron$ absorption factor.
The dereddening corrections are performed by
using the interstellar extinction low of $A_\lambda/A_V=$ 0.02 
($\lambda$=7.8 $\micron$) and 0.031 ($\lambda$=12.4 $\micron$), 
interpolated from table 3 of \citet{rie85}.
The ratio of $A_V/ \tau_{Si}$ = 16.6 \citep{rie85} was also used,
where $\tau_{Si}$=$\tau_{9.7}+0.03$, 
calculated from table 3 of \citet{dr85} 
with the transmission curve of the COMICS 9.7 $\micron$ filter.

\item The first-order color temperature is approximated from 
the corrected 7.8 $\micron$/12.4 $\micron$ brightness ratio,
assuming simple blackbody spectra
[dust emissivity $\epsilon(\lambda)$=constant)].

\item The corrected absorption factor was calculated,
as the ratio of the 9.7 $\micron$ brightness to 
the continuum, interpolated 
using the blackbody curve based on the first-order color temperature.

\item The 7.8 $\micron$ and the 12.4 $\micron$ brightness were again 
corrected by using the corrected absorption factor.

\item The final color temperature was calculated from the corrected
7.8 $\micron$/12.4 $\micron$ ratio.

\end{enumerate}

\begin{figure*} 
	\begin{center} 
		\FigureFile(120mm,120mm){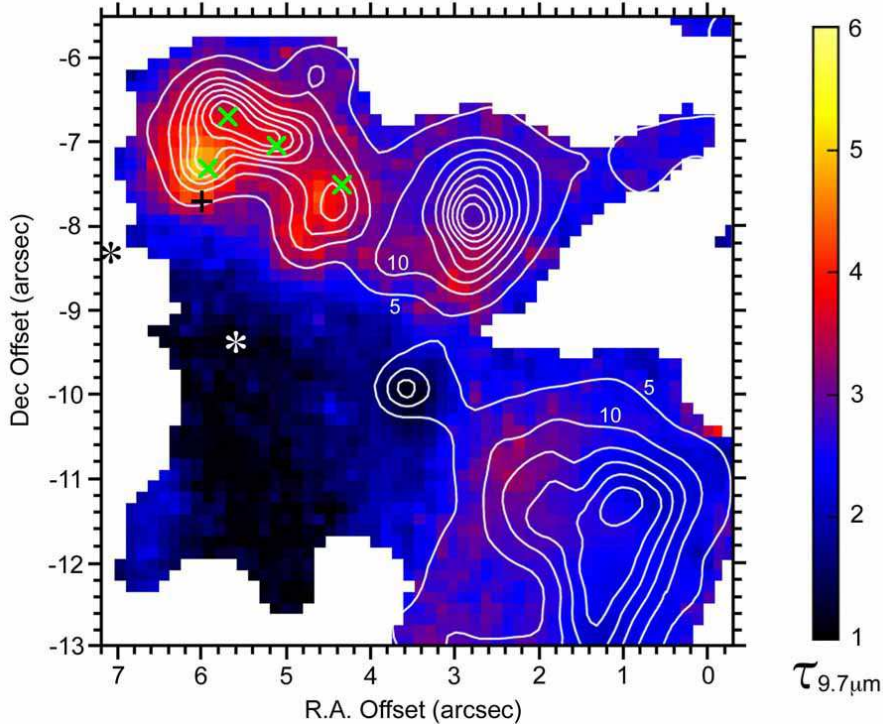} 
	\end{center} 
	\caption{9.7 $\micron$ color-coded optical depth ($\tau_{9.7}$)
	image of the \timeform{7.5"}$\times$\timeform{7.5"} field 
	around IRc2.
	The color image is masked in the area of low S/N ratio
	(9.7 $\micron$ flux density $\leq$ 0.15 Jy/arcsec$^{2}$).
	The overlaid contour shows the 12.4 $\micron$ surface brightness,
	in steps of 5 Jy/arcsec$^{2}$.	
	Four green crosses mark the positions of the 3.8 $\micron$
	IRc2-A, B, D, C, from left to right \citep{dou93}.
	A plus marks the position of radio source I \citep{men95},
	corrected by referring to the proper motions 
	presented by \citet{gom05}.
	A black asterisk shows the position of the ``hot core'' 
	\citep{pla82,wri92,bla96,wil00} and a white asterisk denotes 
	the 865	$\micron$ submillimeter continuum peak, 
	SMA1 \citep{beu04}.
	Symbol size indicates the uncertainty in its position.
	The coordinates are centered on BN.}
	\label{tau} 
\end{figure*}

Figure \ref{tau} shows the 9.7 $\micron$ optical depth
 ($\tau_{9.7}$) distribution around IRc2. 
The error in the $\tau_{9.7}$ is estimated to be a maximum of 0.2.
The average depth of IRc2 
[\timeform{2.5"}$\times$\timeform{2.0"} rectangular area 
centered on (\timeform{5.4"}, \timeform{-7.2"})] is 3.8.
We can see strong absorption in the vicinity of 
the 3.8 $\micron$ IRc2-A \citep{dou93}.
The peak of absorption ($\tau_{9.7}\sim5.5$) 
occurs \timeform{0.1"} east of IRc2-A;
their positional separation is within our astrometric uncertainty.

%%%%% Figure 3 %%%%%

\begin{figure*} 
	\begin{center} 
		\FigureFile(120mm,120mm){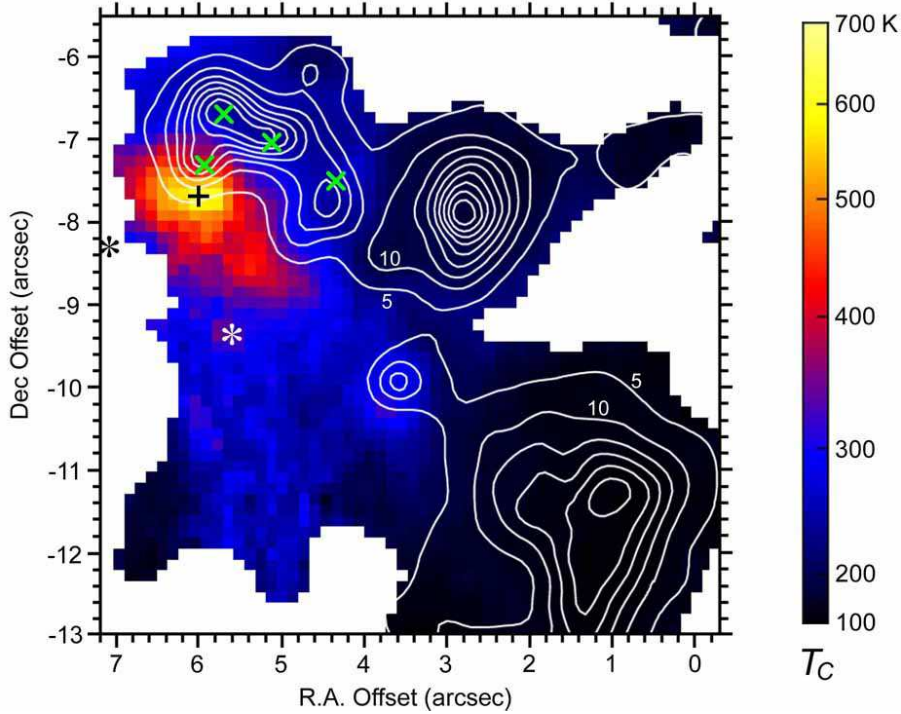} 
	\end{center} 
	\caption{7.8 $\micron$/12.4 $\micron$ color temperature 
	distribution image of the \timeform{7.5"}$\times$\timeform{7.5"} 
	field around IRc2. 
	The color image was omitted in the area of low S/N ratio.
	Contours, symbols, and coordinates are the same as in figure \ref{tau}.
	}
	\label{color_temp} 
\end{figure*}

Figure \ref{color_temp} shows the 7.8 $\micron$/12.4 $\micron$ 
color temperature distribution around IRc2. 
The uncertainty in the corrected 7.8 $\micron$/12.4 $\micron$ ratio 
is estimated at a maximum of 16 \%.
It corresponds to the error of 10 K, 40 K, and 90 K 
at the color temperature of 200 K, 400 K, and 600 K.
For the 12.4 $\micron$ IRc2 main-peak
(\timeform{5.7"}, \timeform{-6.7"}), color temperature is 250 K,  
in agreement with an earlier result \citep{gez98},
while the maximum color temperature, 650 K, 
occurs $\sim$ \timeform{1"} south of the 12.4 $\micron$ IRc2 peak. 
The color temperature peak coincides with the position of source I 
within our astrometric error.
We found another color temperature peak of 350 K 
at (\timeform{5.6"}, \timeform{-9.4"}),
\timeform{1.9"} south-southwest of source I, 
corresponding to the 865 $\micron$ 
submillimeter continuum peak, SMA1 \citep{beu04}.

%%%%% Figure 4 %%%%%

\subsection{Photometry}

The mid-infrared flux densities of IRc2 and other sources were calculated
from dereddened images.
They are summarized in table \ref{tabphoto},
together with the near-infrared 4.0 $\micron$ flux densities 
by \citet{dou93}.
A dereddening correction was computed using the interstellar extinction
law of \citet{rie85} and the dust grain optical constants 
of \citet{dr85}.
The flux densities were integrated over the rectangles listed
in the third column in table \ref{tabphoto}, while
taking the flux density at (\timeform{5.5"}, \timeform{-10.0"}) as 
a background flux density for each wavelength. 
The typical photometric error is estimated to be approximately $\pm$ 5 \%.
The error is over 20 \% in some cases, 
because of the low surface brightness,
or contamination due to background plateau emission.

%%%%% Table 3 %%%%%
\begin{table*}
 \begin{center}
 \caption{Dereddened Total Flux Density 
 for each IR source and radio source I}
\label{tabphoto}
   \begin{tabular}{lcccccccccccc}
   \hline
   &&Photometric&&&
   \multicolumn{6}{c}{Total Flux Density
   \footnotemark[$\ddagger$] (Jy)}\\
   \cline{4-13}\
   &Position\footnotemark[$*$]& 
   Aperture\footnotemark[$\dagger$]
   &4.0\footnotemark[$\S$]&7.8&8.8
   &9.7 &10.5 &11.7&12.4 &18.5&20.8&24.8\\
   Source &(arcsec)& (arcsec) & ($\mu$m)  
   & ($\mu$m) 
   &  ($\mu$m) & ($\mu$m) & ($\mu$m) &  ($\mu$m) 
   & ($\mu$m) &  ($\mu$m) &  ($\mu$m) &  ($\mu$m) \\
   \hline
   	BN &(0, 0)& 3.4$\times$3.4&220 & 420 & 670 & 640 & 630
	   & 740   & 700 & \em960 & \em1500 & \em1500 \\
	Source n &(3.6, --9.9)& 1.4$\times$1.4 & 3.5 &6.8&8.6&13&13&17
	   & 25 & $<$\em70 & $<$\em140 & $<$\em250\\
	IRc2-A &(6.0, --7.3)& 0.7$\times$0.7 & 4.2 & 63 & 65 & 84 & 66 
	   & 78 & 110 & \em130 & \em180 & \em210\\
	IRc2-B &(5.8, --6.7)& 0.7$\times$0.7 & 1.8& 28 & 55& 49 & 49 
	   & 79 & 100 & \em120 & \em180 & \em210\\
	IRc2-C &(4.5, --7.8)& 0.7$\times$0.7 & 1.9 & 16 & 20 & 28 & 32 
	   & 47 & 61 & \em110 & \em170 & \em200\\
	IRc2-D &(5.3, --6.9)& 0.7$\times$0.7 & 0.6& 22 & 46 & 39 & 40 
	   & 66 & 86 & \em120 & \em190 & \em220\\
	IRc2 (Total) &  & &8.5 & 130& 190 & 200 & 190 
	   & 270 & 360 & \em480 & \em700 & \em830\\
	IRc3 &(--1.7, --7.6)& 4.7$\times$5.7 & 2.6 & \em12 
	   & \em28 & \em52 & \em110 & \em220 & \em310 
	   & \em2300 & \ 4100 & \em5400\\
	IRc4 &(1.1, --11.4)& 4.3$\times$4.0 &$\cdots$& \em30 & 45& 100
	  & 180 & 380 & 530 & 2500 & 4200	& 5800\\
	IRc5 &(0.6, --15.0)& 2.5$\times$3.3 &$\cdots$& \em2.7 & \em7.1
	  & \em14 & \em33 & \em65 & \em92	& 750 & 1400 & 1900\\
	IRc7 &(2.8, --7.9)& 3.4$\times$2.5 &4.9 & 42 & \em60 & \em94
      & 130 & 230 	& 320 & \em1000 & 1700 & 2300\\
	IRc11 &(8.9, --7.6)& 3.5$\times$2.3& $\cdots$ & \em12 & \em11 
	  & \em25 &\em28 & 40 & 67 & \em260 & \em510 & 830 \\
	 IRc12 &(10.5, --8.0) & 2.0$\times$1.7 & $\cdots$ & \em7.8 
	   & \em8.5 & \em14 & \em17 & \em21 & 29 & \em100 
	   & \em180 & \em280 \\
	  Source I &(6.0, --7.7)\footnotemark[$\|$] 
	  &0.7$\times$0.7 &$<$\em0.7 &\em11&\em5.5 & \em11 
	  &\em9.6 &\em7.8 &\em11 &\em21 &\em30 &\em40\\
	\hline
	\multicolumn{13}{@{}l@{}}{\hbox to 0pt{\parbox{180mm}
	{\footnotesize
	\footnotemark[$*$] 12.4 $\micron$ peak position,
	offset from BN 
	($\Delta\alpha$, $\Delta\delta$).\\
	\footnotemark[$\dagger$] Rectangular size 
	[(R.A.) $\times$ (Dec)].\\
	\footnotemark[$\ddagger$] Typical photometric error is estimated 
	to be approximately $\pm$ 5 \%. 
	It is sometimes over $\pm$ 20 \% (in italics), 
	because of low surface brightness and/or 
	the surrounding plateau emission.
	At $\lambda>$ 18 $\micron$, source n is not distinct from 
	its surrounding extended emission.\\
	\footnotemark[$\S$] The flux densities at 4.0 $\micron$ are 
	taken from \citet{dou93}, followed by the dereddening corrections 
    in the procedures described in \ref{sec:ext}.\\
	\footnotemark[$\|$] Radio position \citep{men95},
	corrected by the proper motion presented by \citet{gom05}.
	}\hss}}
   \end{tabular}
 \end{center}
 \end{table*}

\subsection{Spectroscopy}

\begin{figure*} 
	\begin{center} 
		\FigureFile(100mm,60mm){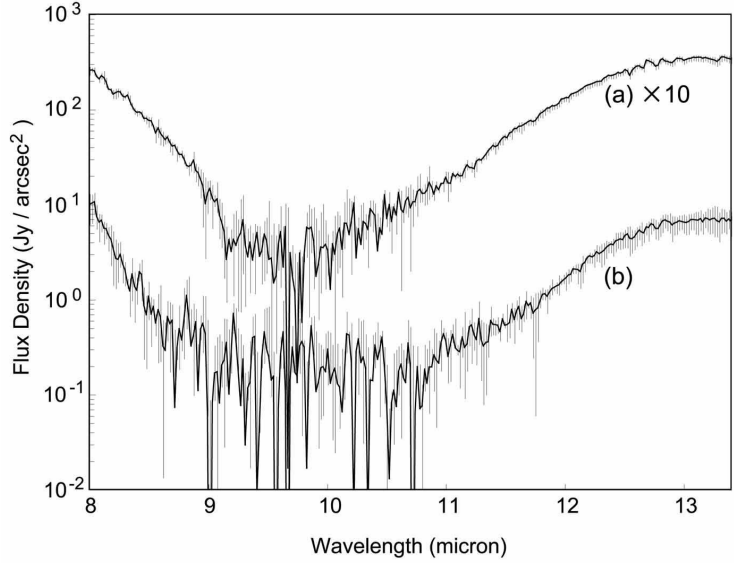} 
	\end{center} 
	\caption{$N$-band low resolution spectra of IRc2,
		at the position of 12.4 $\micron$ main-peak (a) and source I (b),
		shown in figure \ref{color_sp}.}
	\label{spectroscopy} 
\end{figure*}

\begin{figure*} 
	\begin{center} 
		\FigureFile(140mm,95mm){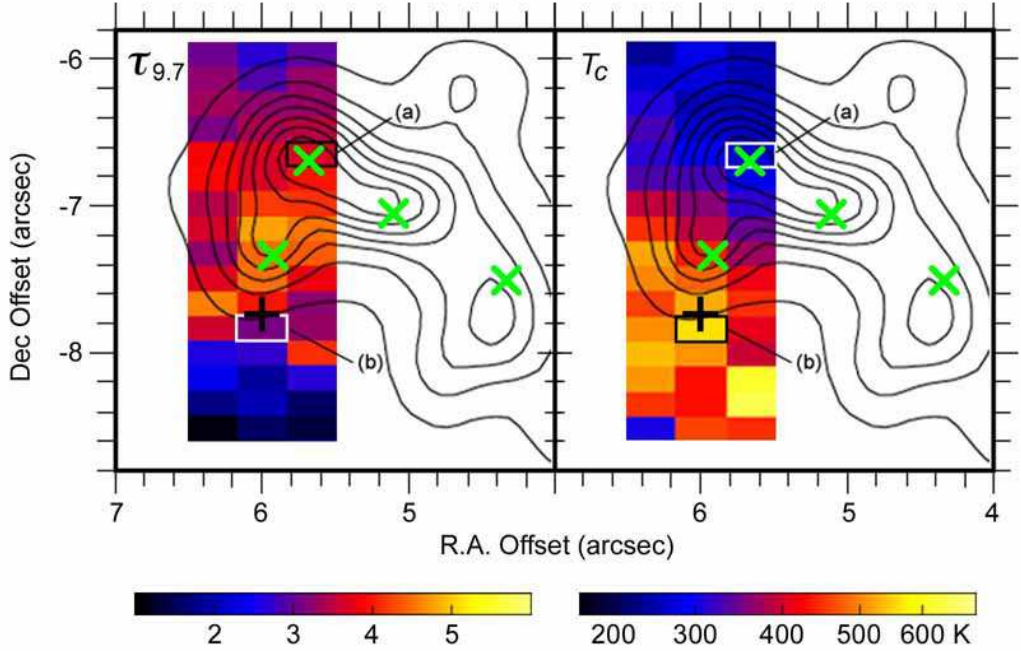} 
	\end{center} 
	\caption{9.7 $\micron$ optical depth (left) and the
	7.8 $\micron$/12.4 $\micron$ color temperature (right) images
	of IRc2, constructed from the spectra.
	Overlaid  contours and symbols are the same as in figure \ref{tau}.
		The upper and lower spectra in figure \ref{spectroscopy} are obtained 
		at the positions of (a) and (b), respectively.}
	\label{color_sp} 
\end{figure*}

Figure \ref{spectroscopy} presents the $N$-band low resolution
spectra of IRc2, obtained at the positions of the 
12.4 $\micron$ main-peak and radio source I.
Except the silicate absorption,
the spectra show no strong atomic or broad feature,
such as the 8.6, 11.2, and 12.7 $\micron$ 
polycyclic aromatic hydrocarbon (PAH) features (e.g., \cite{kas06}),
which would have affected the discussion in section \ref{sec:ext}.
Figure \ref{color_sp} shows 
the 9.7 $\micron$ optical depth image (left)
and the 7.8 $\micron$/12.4 $\micron$ color temperature 
image (right) created from the spectra,
through the same procedures discussed in section \ref{sec:ext}.
The results are in agreement with figures \ref{tau}
and \ref{color_temp}.

%%%%% Figure 5 %%%%%

%%%%% Figure 6 %%%%%

\section{Discussion}
\subsection{Background of the discussion about the relation
between IRc2 and source I}

The idea that IRc2 is a dominant energy source for the KL nebula
was supported by early observations 
showing IRc2 to be associated with source I,
within the positional errors at that time.
Since IRc2 was found to be offset from radio source I,
we have come to have many new discussions about
the physical relation between IRc2 and source I.
\citet{gez98} revealed that the total 
infrared luminosity of IRc2 is only $L\sim$1000$\LO$, 
and concluded that IRc2 is a dense dust cloud heated internally 
by the near-infrared sources found by \citet{dou93} 
and also externally by the luminous star at source I.
They pointed out  that a silicate absorption peak
and a color temperature peak both occur near IRc2,
whereas the detailed structure of the distribution of absorption 
and color temperature in the vicinity of IRc2 was not fully 
resolved from their \timeform{1.1''}-FWHM resolution data,
because IRc2 is in close proximity ($\sim\timeform{1''}$) of  source I.
For example, their maximum silicate absorption factor 
(9.8 $\micron$ line-to-continuum ratio), 52,
is reproduced by the average optical depth of IRc2 in our result.
\citet{gre04} speculated that knots in IRc2 as well as source I 
are individual sites of star formation.
In support of this speculation, they noted that 
the relative distributions of emission in IRc2
from 2.2 $\micron$ to 22 $\micron$
do not exhibit an overall systematic pattern,
such as central heating by one or two dominant energy sources.
They also mentioned that from among all knots in IRc2, 
only IRc2-C is associated with a cluster of OH masers 
and X-ray emission.
\citet{shu04} pointed out that the near-infrared IRc2 knots
all appear on the periphery of the 12.5 $\micron$ emission.
They also pointed out that 
the volume-averaged density for IRc2 would be much larger than that of 
a typical star-forming cloud core,
if IRc2 knots were all embedded protostars. 
For these reasons,
they suggested the possibility that IRc2 is illuminated and heated
by source I,
whereas they also suggested another possibility 
that both external and internal heating
play a role in the emission from IRc2, 
as would be expected if a low mass, dense protostar cocoon are
placed close to an energy source.
\citet{rob05} resolved a conspicuous point source at the location of
the IRc2-A knot in their dereddened mid-infrared images.
They considered a possibility that source I and IRc2-A are
representing deeply embedded individual sources, and that
other IRc2 knots are diffuse sources heated externally by IRc2-A.

\subsection{Dereddened spectral energy distributions of IRc2
and other sources}
\label{sec:SED}

\begin{figure*}
	\begin{center} 
		\FigureFile(160mm,70mm){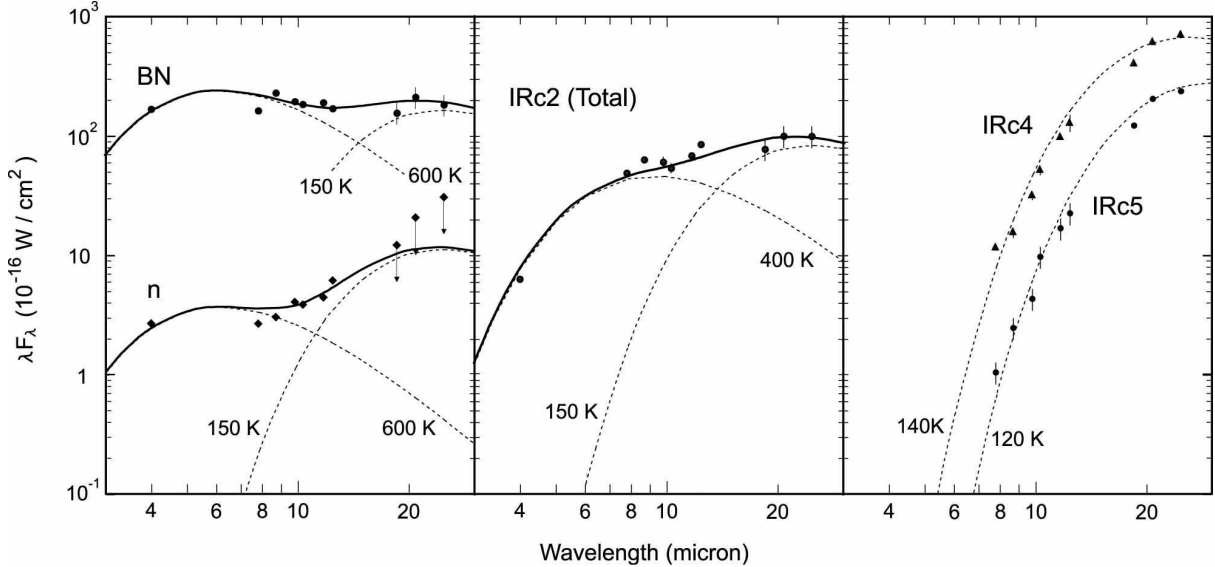} 
	\end{center} 
	\caption{Dereddened spectral energy distributions (SEDs) 
	of BN and  source n (left panel), IRc2 (middle panel), 
	and IRc4 and IRc5 (right panel). 
	Dashed curves show single temperature
	Planck functions.
	Solid curves show the sum of the dashed curves in each panel.
	}
	\label{sed} 
\end{figure*}

Figure \ref{sed} shows the dereddened 4--24.8 $\micron$ 
spectral energy distributions (SEDs) of some sources listed 
in table \ref{tabphoto},
together with the curves of Planck functions fitted to the SEDs.
Although the SED of BN cannot be reproduced 
by a single blackbody emission,
it can be fitted by a two-temperature component model, 
e.g., 600 K and 150 K.
The SED of source n, already identified to be a stellar object 
with a circumstellar disk \citep{shu04,gre04},
is also reproduced by a similar model.
In contrast,
the SEDs of IRc4 and IRc5 can be reproduced by single temperature
components at 140 K and 120 K, respectively.
It is a plausible conclusion that IRc4 and IRc5 are simple dust clouds
heated by an external energy source(s).
The SED of IRc2 cannot be reproduced by a single blackbody model;
a warmer component (e.g., 400 K) seems to overlap 
a cooler component (e.g., 150 K), similar to that of BN or source n.
In its appearance, IRc2 seems to be a self-luminous object.
In the following section, however,
we describe the origin of the ``warmer component'' in the SED of IRc2,
which shows that IRc2 is not self-luminous.

%%%%% Figure 7 %%%%%

\subsection{Physical relation of source I to IRc2}

\begin{figure*} 
	\begin{center} 
		\FigureFile(85mm,85mm){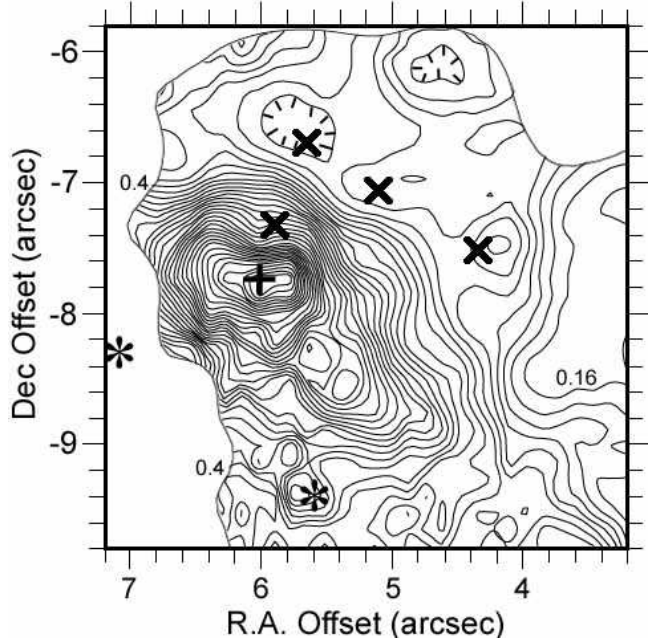} 
	\end{center} 
	\caption{Extinction-corrected 7.8 $\micron$/12.4 $\micron$ 
	brightness ratio contour map of the 
	\timeform{4"}$\times$\timeform{4"} field 
	around IRc2.
	Four crosses mark the positions of the 3.8 $\micron$
	IRc2-A, B, D, C, from left to right.
	A plus and two asterisks mark the positions of the radio source I,
	the hot core, and SMA1, same as in figure \ref{tau}.
	The contour step is 0.03.
	The contour was omitted in the area of low S/N ratio.
	}
	\label{color_cont} 
\end{figure*}

In this subsection, 
we discuss the origin of the ``warmer component'' 
of the emission from IRc2.
The simplest interpretation,
widely accepted for $\sim$ 30yr, 
is the presence of an internal heating source.
In figure \ref{color_temp}, however,
the highest color temperature peak occurs at source I, 
\timeform{1"} south of the 12.4 $\micron$ peak of IRc2,
and a local color temperature distribution has 
increasing gradients from IRc2-B, C, and D 
toward source I.
Figure \ref{color_cont} shows the extinction-corrected
7.8 $\micron$/12.4 $\micron$ brightness ratio contour map. 
It is a color temperature map similar to figure \ref{color_temp}, 
whereas it displays weaker levels to clearify the detailed
structure around IRc2.
The 7.8 $\micron$/12.4 $\micron$ brightness ratios for 
IRc2-B, C, and D are approximately 0.3.
They are only 24 \% of the ratio for source I 
and correspond to the color temperature of $\sim$ 265 K.
A small peak of the color temperature is seen 
near to the westernmost component, IRc2-C, 
but no peak is near IRc2-B and D.
No temperature peak is seen near IRc2-A, 
because it is probably overlapped 
by the shoulder of the large peak at source I.

%%%%% Figure 8 %%%%%

A temperature peak should
indicate the location of a luminosity source,
on the argument that temperature gradients trace
the direction of energy flow.
Therefore, the color temperature distribution suggest 
a luminosity source at the position of source I.
The temperature distribution does not necessarily rule out 
the possibility of other self-luminous sources embedded in 
the location where no temperature peak is seen.
However, if there are other dominant luminosity sources within IRc2,
they must be obscured by the large extinction that makes 
even the structure of the local color temperature peak 
surrounding them invisible.
Therefore, we conclude that no dominant
self-luminous object is embedded in IRc2,
though faint self-luminous sources may be embedded.

As previously presented in subsection \ref{sec:ext},
the 3.8 $\micron$ emission peak, IRc2-A, 
coincides with the absorption peak.
It is surprising that the shorter wavelength emission exhibits 
its intensity peak at the local absorption peak position,
when the 3.8 $\micron$ emission arises from
self-luminous objects deeply embedded in IRc2-A.  
This naturally leads to the interpretation that near-infrared emission
from IRc2-A is due to scattering of radiation from an external source;
source I is the most likely candidate.
Two other near-infrared IRc2 knots, IRc2-B and D, 
also reflect surface layers of dense clouds
illuminated by an external source.
There remains the possibility that a faint self-luminous object
to be embedded in IRc2-C.
However, it is not easy to conclude 
that the object is a dominant energy source of IRc2.

Previous polarimetric observations by \citet{dou93} show that 
IRc2-A and IRc2-B are highly polarized at 3.8 $\micron$
(14\% and 17\%, respectively), with position angles 
practically perpendicular to the direction of source I.
Their results support the suggestion 
that IRc2-A and IRc2-B are illuminated under conditions of single scattering
by source I, but not by IRc2-C.
Low polarization of 8.3 \% 
for IRc2-C \citep{dou93} suggests 
the possibility of IRc2-C is self-luminous.

%%%%% Figure 9 %%%%%
\begin{figure*} 
	\begin{center} 
		\FigureFile(120mm,120mm){figure9.eps} 
	\end{center} 
	\caption{Contour maps of IRc2 and neighboring sources:
	(a) 7.8 $\micron$ contour map.
	The contours start at 2 Jy/arcsec$^{2}$ and continue 
	in steps of 4 Jy/arcsec$^{2}$. 
	(b) 20.4 $\micron$ contour map.
	The contours start at 5 Jy/arcsec$^{2}$, 
	runnning in steps of 10 Jy/arcsec$^{2}$.	
	(c) 5.0 and 4.0 $\micron$ contour map
	from \citet{dou93}.
	(d) 9.7 $\micron$ optical depth map.
	The contour levels are from 2.0 to 5.3 Jy/arcsec$^{2}$
	in steps of 0.3 Jy/arcsec$^{2}$.
	The map is masked in the area of low S/N ratio.
	Symbols are the same as in figure \ref{color_cont}.
	}
	\label{hikaku} 
\end{figure*}

%%%%% Figure 10 %%%%%
\begin{figure*} 
	\begin{center} 
		\FigureFile(160mm,110mm){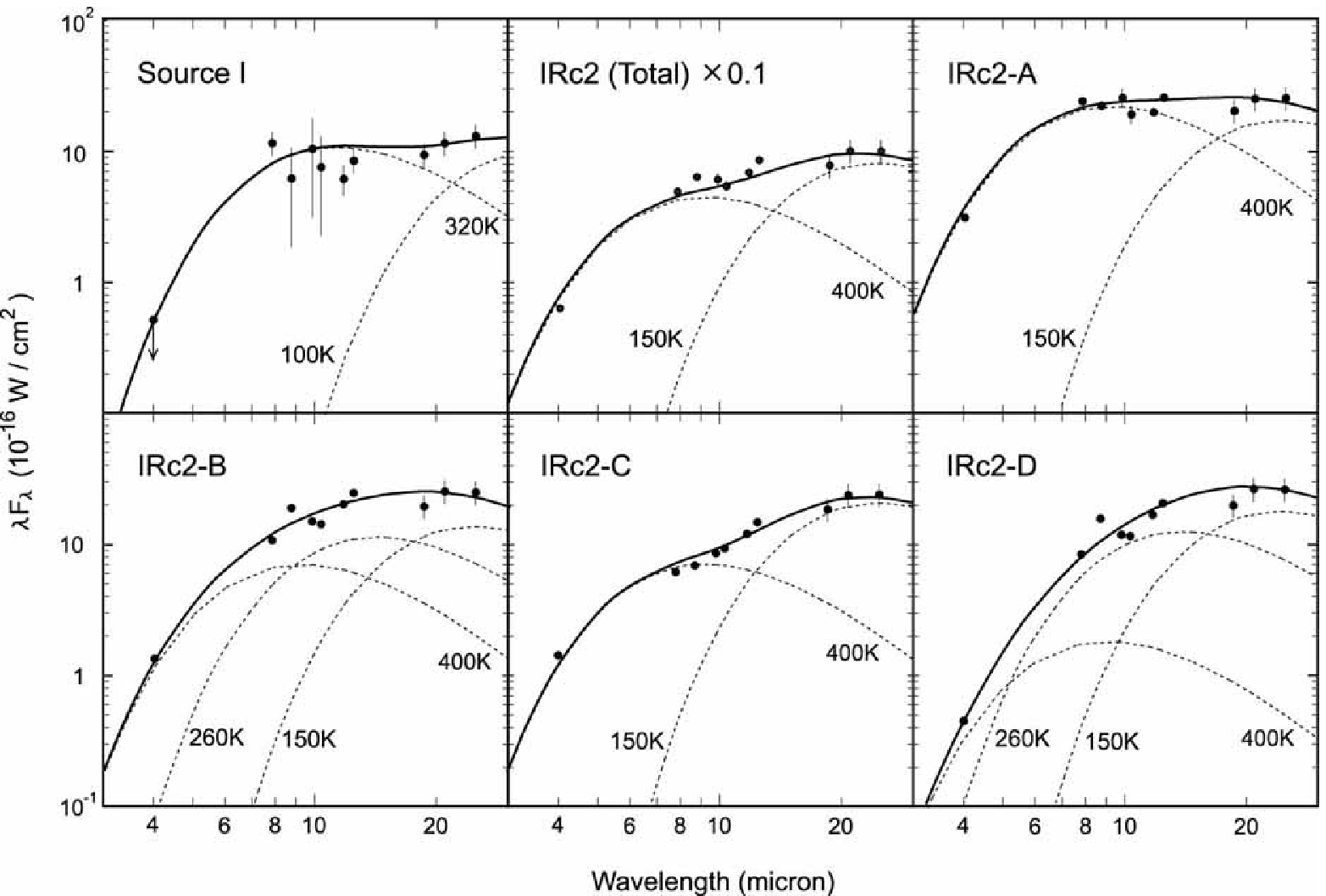} 
	\end{center} 
	\caption{Dereddened spectral energy distributions (SEDs) 
	of source I and individual IRc2 knots.
	Dashed curves show single-temperature
	Planck functions, and
	solid curves show the sum of the dashed curves.
	}
	\label{IRc2sed} 
\end{figure*}

For helping the discussion concerning the origin of 
the warmer component in the SED of IRc2, 
the IRc2 contour maps at 7.8 $\micron$ and
20.4 $\micron$ are presented in figure \ref{hikaku},
together with the near-infrared contour map of \citet{dou93}
and the 9.7 $\micron$ optical depth distribution.
The images at 7.8 $\micron$ and 20.4 $\micron$ roughly
correspond to the spatial distributions of the warmer component 
and the cooler component, respectively.  
In figure \ref{hikaku}, 
the cooler component is dominant at IRc4 and IRc7,
whereas both the warmer component and the cooler component
are found at IRc2, as previously noted in subsection \ref{sec:SED}.
The appearances of IRc2 are different in each
image of 7.8 $\micron$ and 20.4 $\micron$,
which shows a possibility that the sources of the warmer component 
and the cooler component are different. 
The origin of the cooler component is certainly reradiation 
from dust cloud.
It is also the case with IRc4 and IRc5 
as discussed in subsection \ref{sec:SED}.
For IRc2, a morphological correlation is seen 
between the 7.8 $\micron$ image and the near-infrared 
4.0 $\micron$/5.0 $\micron$ image of \citet{dou93}.
In figure \ref{hikaku}, in addition, it is easy to recognize
a close correlation between the 7.8 $\micron$ image 
and the optical depth distribution at IRc2. 

We suggest that the warmer component of the
mid-infrared emission from IRc2 arises from the same source
as the near-infrared emission.
This means that the mid-infrared warmer component emission
from IRc2 is the scattering of radiation from source I.
Then, the warmer component emission should suffer 
not only line-of-sight extinction, but also significant amounts of 
extinction along the path from source I to IRc 2.
This is consistent with a larger optical depth of IRc2,
and also with the morphological coincidence
between the 7.8 $\micron$ image 
and the optical depth distribution of IRc2.

Figure \ref{IRc2sed} shows the dereddened SEDs of 
source I and total and individual IRc2 knots.
The SEDs of IRc2-A and IRc2-C can be reproduced by the combination 
of a warmer component (400 K) and a cooler component (150 K).
Here, we must note that the previously presented 
7.8 $\micron$/12.4 $\micron$ color temperature 
(figures \ref{color_temp}, \ref{color_sp}, and \ref{color_cont}) 
is estimated based on the total amount of each temperature component.
It can be easily recognized that the cooler component represents
emission from the dust cloud heated externally.
We will discuss the origin of the warmer component
based on a suggestion that it is scattered light.
For ``astronomical silicate,'' 
the scattering efficiency generally decreases with wavelength
in the infrared region.
At 8 $\micron<\lambda<$ 10 $\micron$, 
however, the scattering efficiency increases with wavelength;
the efficiency at 12.4 $\micron$ is 1.7 times larger than 
that at 7.8 $\micron$ \citep{dr84,dr85}.
Therefore, scattering process brings a reddening effect 
to the SED at this wavelength.
It leads to a decrease in 
the 7.8 $\micron$/12.4 $\micron$ color temperature;
the 7.8 $\micron$/12.4 $\micron$ brightness ratio of 1.10
for the unit of W cm$^{-2}$, 
which corresponds to the color temperature of 400 K,
can be explained by scattering of the radiation
with the original 7.8 $\micron$/12.4 $\micron$ ratio of 1.87,
which indicates a color temperature of 600 K.
Therefore, the 400 K warmer component in the SED  
can be explained by the result of the scattering of 
600 K blackbody radiation in this wavelength region.
IRc2-B and IRc2-D cannot be reproduced  
by a two-component blackbody model.
Another blackbody component (e.g., 260 K) seems to be overlapped
on the two-component model, though it is of unclear origin.
There are some possibilities; 
a multiple scattering effect, the presence of hot dust, 
or the presence of deeply embedded sources.

Though the SED of source I cannot be easily fitted by a single- or 
a two-temperature component model, 
it is roughly reproduced by a combination of a 320 K component 
and  a 100 K component.
The SED probably represents the radiation from hot dust 
heated by source I, but not from (embedded) source I, itself.
Therefore, the extinction toward source I is estimated to be 
smaller than that toward IRc2.
Direct emission from source I is concealed by dense dust, 
and in the mid-infrared region, 
it is visible only as the scattered light from IRc2.

\begin{figure*} 
	\begin{center} 
		\FigureFile(160mm,60mm){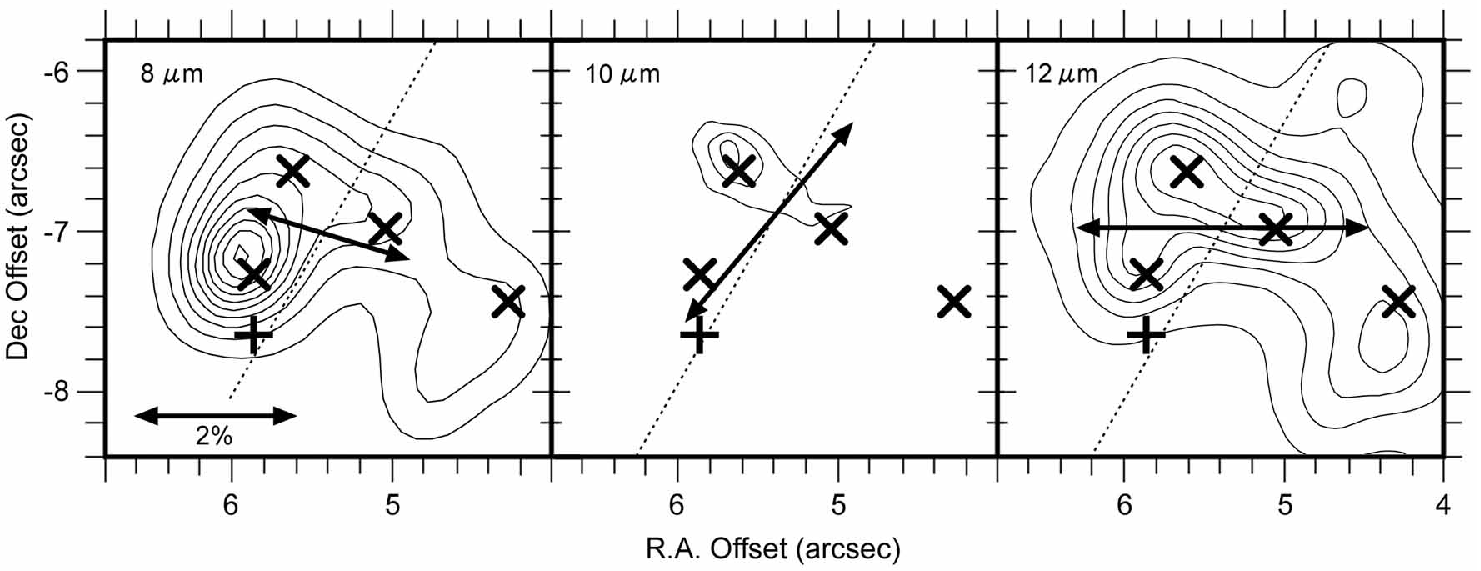} 
	\end{center} 
	\caption{Polarization vectors of IRc2 
	at 8 $\micron$ (left), 10 $\micron$ (middle), 
	and 12 $\micron$ (right), obtained in 1990 
	with an aperture of \timeform{2.9''} diameter \citep{smi00}.
	Dotted lines show the overlying magnetic field direction \citep{ait97}.	
	Overlaid contour maps represent the undereddened surface brightness
	at 7.8 $\micron$ (left), 9.7 $\micron$ (middle), 
	and 12.4 $\micron$ (right), respectively.
	The contour levels are from 2 to 22 Jy/arcsec$^{2}$
	in steps of 4 Jy/arcsec$^{2}$ (left),
	0.5, 0.6, 0.7 Jy/arcsec$^{2}$ (middle),
	and from 5 to 40 Jy/arcsec$^{2}$ 
	in steps of 5 Jy/arcsec$^{2}$ (right).
	Four crosses and a plus mark the positions of the 3.8 $\micron$
	IRc2 knots and source I,
	same as in figure \ref{color_sp} or figure \ref{color_cont}.
	They are corrected to those in the epoch of 1990, 
	referring to the proper motions by \citet{gom05}.
	}
	\label{fig11} 
\end{figure*}

Mid-infrared spectropolarimetric observations of IRc2 were 
reported by \citet{ait93}, \citet{ait97}, and \citet{smi00}.
Their observations, with an aperture of \timeform{2.9''} diameter, 
showed that the polarization position angles of IRc2 
at 8 $\micron$, 10 $\micron$, and 12 $\micron$ were 
73$\pm4\degree$, 130$\pm20\degree$, and 90$\pm2\degree$, 
respectively \citep{smi00}.
They proposed a polarization mechanism of a two-component model 
due to dichroic absorption and emission processes by aligned grains,
to explain variations of the polarization fraction and position angle.
They argued that the observed polarization of IRc2 was 
reproduced by combining  
the absorptive component with a position angle of 120$\degree$
and the emissive component with 60$\degree$ \citep{ait97}.
Figure \ref{fig11} shows the mid-infrared polarization and 
the undereddened surface brightness of IRc2. 
At 8 $\micron$ and 12 $\micron$, we infer that
the scattering effect plays a considerable role in 
the polarization mechanism of IRc2, whereas 
the 10 $\micron$ polarization is thought to be predominantly
due to dichroic absorption showing the direction of an 
overlying magnetic field of $\sim$120$\degree$ \citep{ait97},
because (presumed) scattered emission is too weak 
to contribute to the polarization mechanism.

%%%%% Figure 11 %%%%%

We conclude that the mid-infrared emission from IRc2,
especially the warmer component in the SED,
is scattered light that originally arises from source I,
because (1) no dominant color temperature peak is seen at IRc2,
(2) the near-infrared emission from IRc2 is most likely to be 
scattered radiation because of large polarization,
and also a good correlation is seen in the appearance of IRc2 
between the mid-infrared warmer component 
and the near-infrared emission,
(3) a good correlation between the distribution of optical depth and 
the appearance of the warmer component is seen at IRc2, and
(4) the warmer component can be reproduced by reddening process 
in scattering of emission from source I.
Therefore, IRc2 itself, is not a dominant luminosity source,
and also not a dominant energy source of KL, 
though some faint self-luminous objects may be embedded.

\subsection{Comparison with recent radio observations}

Recent millimeter observations revealed the existence of 
a ``hot core,'' with an estimated temperature of 350 K
 \citep{beu05}, \timeform{1.2"} southeast from source I.
However, no mid-infrared counterpart to the hot core has been detected,
probably due to high extinction.
In contrast to the hot core,
the submillimeter continuum peak SMA1 is detected as 
a color temperature peak
in figures \ref{color_temp} and \ref{color_cont},
though we cannot identify any 
counterpart to SMA1 at each wavelength image.
SMA1 is a protostellar object
concealed by dense dust,  just like source I.

\citet{rei07} and \citet{mat10} proposed that 
source I has an edge-on disk whose rotation axis is oriented along
northeast-southwest (NE-SW).
The newly proposed NE-SW axis is different from the   
northwest-southeast (NW-SE) axis
previously proposed in 1980's \citep{has84}.
As \citet{tes10} pointed out, the NE-SW axis disk model is 
hard to reconcile with the observed NW-SE direction of 
the large-scale outflow cavity (e.g., \cite{eri82}),
under the hypothesis that the disk is responsible for the bipolar outflow 
that blows in the direction of the axis of the disk.
Our model, that IRc2 represents 
the scattering of radiation from source I, 
also seems to be inconsistent with the new NE-SW axis disk model,
because IRc2 is approximately located northwest of source I.
However, the inconsistency is canceled
when the dominant illuminator of IRc2 is 
the rotating disk around source I
rather than the photosphere of its central star,
while the surrounding near-infrared nebulosity is due to reflection 
of the light from the photosphere \citep{mor98,tes10}.

As noted in subsection \ref{sec:astrometry},
the proper motions of BN, source I, and source n 
were reported by many authors
and the dynamical history of the Orion BN/KL region was also discussed.
There are claims that the large-scale high-velocity outflow
is not the typical one powered by a disk accretion.
\citet{gom08} proposed that a dynamical decay of a 
protostellar cluster resulted in the formation of a tight binary
(source I) and the high-velocity ejection of BN and source n.
Then, \citet{bal11} proposed 
that the gravitational potential energy
released by the formation of the source I binary resulted in
the stellar ejection, and also powered large-scale outflow.
\citet{god11} presented  another scenario that source I 
originally existed as a softer binary 
and that the gravitational energy was released 
in the process of hardening of the softer binary,
after a dynamical interaction between source I and BN.
The proper motion study of IRc2, e.g., relative to source I, 
will be helpful information  
for discussing the evolutional history of source I and
the origin of the high-velocity outflow.

\section{Conclusion}

In this paper we present mid-infrared imaging and spectroscopic
data of the Orion BN/KL region.
The distributions of silicate absorption strength and 
color temperature were calculated around IRc2 
with our \timeform{0.4''}-resolution data.
The detailed structure of the color temperature distribution around IRc2,
including source I and SMA1, has been revealed in the mid-infrared
region for the first time. 
In the vicinity of IRc2,
the peak color temperature occurs at source I,
and an increasing temperature gradient is seen 
from IRc2 toward source I.
The warmer component of the mid-infrared emission from IRc2 
can be reproduced by the scattering of radiation from source I.
IRc2, itself, is not a self-luminous object,
but illuminated and heated by an external source,
an embedded young stellar object located at source I.

\vspace{1pc}\par
We would like to thank all the staff members of 
the Subaru telescope and the staff of the SMOKA science archive system.
We are also very grateful to an anonymous referee for detailed comments 
that improved the quality of our original manuscript.
Data analysis were in part carried out on the computer system
at the Astronomy Data Center, 
National Astronomical Observatory of Japan.

\end{document}